\documentclass[reprint,showkeys]{revtex4-2}
\pdfoutput=1
\usepackage{graphicx}
\usepackage{bm}
\usepackage{amsmath}
\usepackage{xcolor}
\usepackage[colorlinks=true,linkcolor=blue,urlcolor=blue,citecolor=blue]{hyperref}

\newcommand{\rcm}{\mbox{cm$^{-1}$}}
\newcommand{\SSst}{$^1\Sigma^{+}_{\mathrm{u}}$}
\newcommand{\SPst}{$^1\Pi_{\mathrm{u}}$}
\newcommand{\TSst}{$^3\Sigma^{+}_{\mathrm{u}}$}
\newcommand{\TPst}{$^3\Pi_{\mathrm{u}}$}

\begin{document}

\title{On the 1$^1\Sigma^{+}_{\mathrm{u}}$ electronic state in strontium dimer}

\author{Jacek Szczepkowski}
\email{jszczep@ifpan.edu.pl}
\author{Wlodzimierz Jastrzebski}%
\email{jastr@ifpan.edu.pl}
\author{Anna Grochola}
\affiliation{Institute of Physics, Polish Academy of Sciences,
	al.~Lotnik\'{o}w~32/46, 02-668~Warsaw, Poland}
\author{Pawel Kowalczyk }
\email{Pawel.Kowalczyk@fuw.edu.pl}
\affiliation{Institute of Experimental Physics, Faculty of Physics,
	University of Warsaw, ul.~Pasteura~5, 02-093~Warszawa, Poland}%

\begin{abstract}
The laser technique of polarisation labelling of levels has been used to study the excited 1$^1\Sigma^{+}_{\mathrm{u}}$ state of $^{88}$Sr$_2$, in most part heavily perturbed by the neighbouring electronic states. By observation of completely resolved rovibronic spectra we have determined positions of 2556 spectral lines originating from the labelled levels in the ground X$^1\Sigma^{+}_{\mathrm{g}}$ state and terminating on levels of the perturbed complex. The experimental results are confronted with the existing theoretical predictions.
\end{abstract}

\date{\today}

\maketitle

\section{Introduction}

Despite recent interest in ultracold strontium dimers \cite{Skomorowski1,Stellmer,Reinaudi,Ciamei1,Kondov,Leung} relatively little is known about the structure of excited electronic states of this molecule. Experimental studies are hindered by congestion of spectra due to the presence of several isotopologues of Sr$_2$ at relatively high abundance as well as strong interaction between singlet and triplet states, which makes both manifolds observable and causes formation of irregular spectral patterns. As a result, beside the well known ground X$^1\Sigma^{+}_{\mathrm{g}}$ state, only lower parts of the 1\SSst, 1\SPst\ and 2\SSst\ states are well characterised by state of art spectroscopic methods \cite{Stein1,Stein2,Stein3}. Theoretical investigation on electronic structure of Sr$_2$ also meets with difficulties related in particular to a large number of electrons (76, four of them being valence electrons) with highly correlated behaviour. The theoretical papers known to us \cite{Allouche, Czuchaj, Kotochigova, Skomorowski2, my_Sr2} provide often contradictory results and comparison with experiments shows that the accuracy of these calculations is questionable.

\begin{figure}
	\includegraphics[width=.95\linewidth]{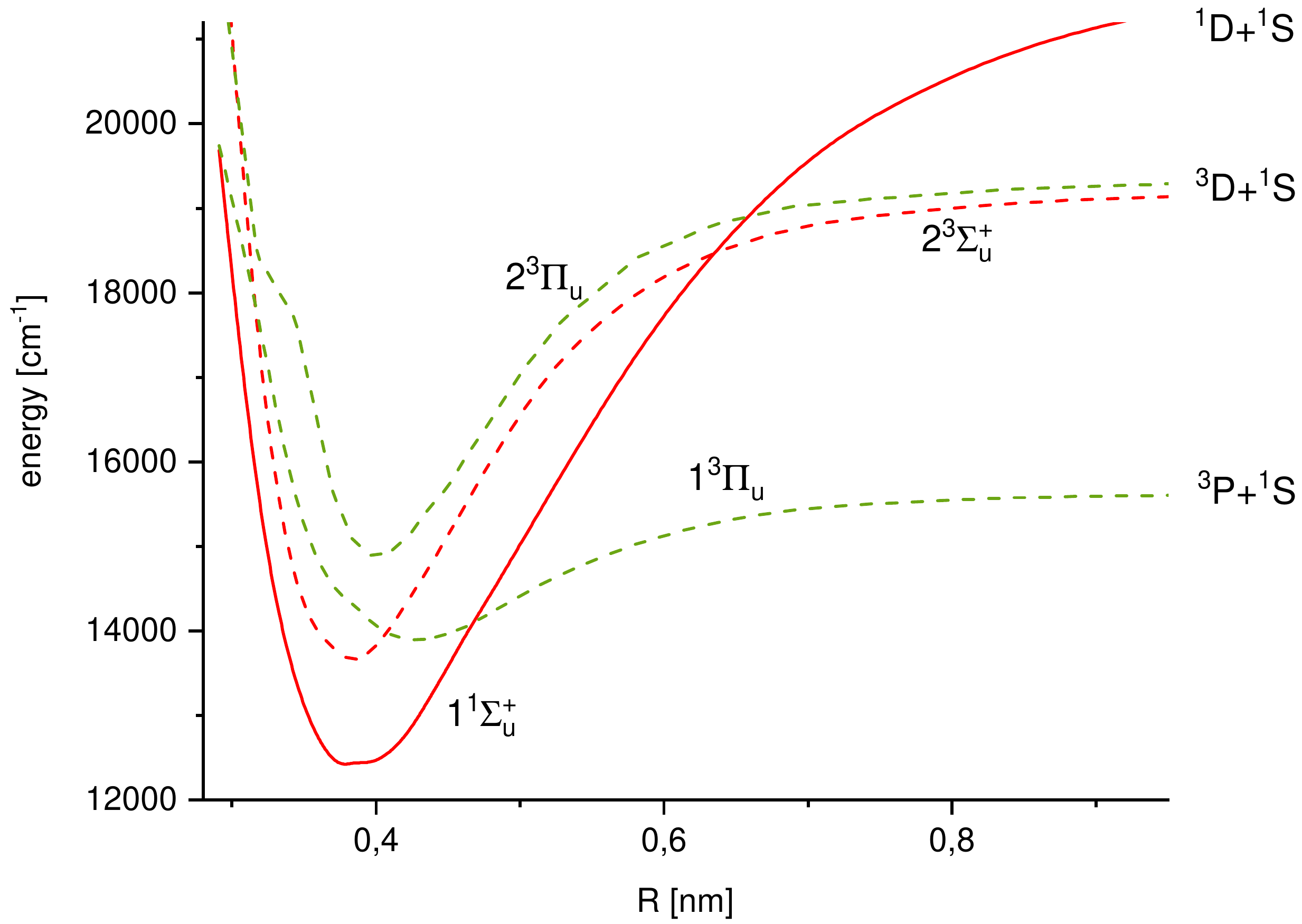}
	\caption{Potential energy curves of Sr$_2$ relevant to this paper, calculated in Hund's case (a) by Boutassetta \textit{et al.} \cite{Allouche,Allouche2}.}
	\label{Fig1}
\end{figure}

In the present paper we investigate experimentally the lowest excited state of $^{88}$Sr$_2$ available by optical transition from the ground state, i.e. the (A)1\SSst\ state. This state was observed before by Stein \textit{et al.} \cite{Stein3} but only the lowest vibrational levels up to $v’=10$ were characterised. The gross properties of the electronic states of strontium dimer, located in the studied energy range, can be inferred from theoretical calculations. All of them predict a presence of four states there, besides the 1\SSst\ also 2\TSst, 1\TPst\ and 2\TPst, although their mutual positions vary depending on different approaches (see Figure~\ref{Fig1} for an example of the relevant states calculated by Boutassetta \textit{et al.} \cite{Allouche,Allouche2}). It is important that all four states can be mutually coupled via spin-orbit or rotational interactions. In our experiment we employed the optical-optical double resonance polarisation labelling spectroscopy technique which proved successful in studies of many excited states in diatomic molecules. In the following sections we present a principle of the method together with experimental details, analysis of the recorded spectra and discussion of our findings in view of the existing theoretical predictions.

\section{Experimental method and apparatus}
\label{Exp_method}

The V-type double resonance polarisation labelling spectroscopy (PLS) has been described in the literature (see for example \cite{Teets1, Grochola1}) and extensively used for studying excited electronic states of simple molecules \cite{Teets2,Ehret,Niay,WJ1,Pashov1,Bang1,Bang2,WJ2,Pashov2}. In our version of the method we use beams of two lasers. A fixed frequency, weak probe laser beam is tuned to resonance with a known rovibronic transition in the investigated molecule, thus selecting ('labelling') the lower level in the involved transition. A tuneable, polarised strong beam of the second laser establishes by means of optical pumping a difference in population among $M$ sublevels of the initial (and final) levels of the pump transition. The molecular sample is placed between two crossed polarisers inserted in the path of the probe beam. A linearly polarised probe beam traversing the prepared sample is therefore stopped by the second, crossed polariser at all frequencies except those at which the sample is anisotropic. Accordingly, the signal recorded behind the second polariser corresponds only to frequencies of transitions originating from the labelled level in the ground state. (Possible transitions from or to the upper labelled level can be distinguished by their spectral features and actually have not been observed in the present experiment.) 

In practical realisation of the above idea in our experiment the pump light was generated by a parametric oscillator/amplifier system (OPO/OPA, Sunlite Ex, Continuum) pumped with the third harmonic of an injection seeded Nd:YAG laser (Powerlite 8000). Depending on the spectral region of interest either the idler or the signal beam from the OPO/OPA system was used, for the purpose of the present experiment tuneable between 645 and 770~nm. The system delivered pulses of 10~ns duration, spectral width 0.1~\rcm\ and energy of few tens of mJ, however, to avoid saturation of molecular signals, the laser intensity was attenuated with neutral filters of variable optical density. Frequency measurements were referenced to the optogalvanic lines of an Ar hollow cathode lamp and fringe markers of a 0.5~cm long Fabry-P\'{e}rot interferometer, providing accuracy of measured laser wavenumbers estimated as 0.1~\rcm. As a source of the probe (labelling) light we used a cw single-mode ring dye laser (Coherent 699-21 operated on Rhodamine 6G dye), with a wavelength controlled using a HighFinesse WS-7 wavemeter, set at fixed frequencies of selected transitions from the X$^1\Sigma^{+}_{\mathrm{g}}$ ground state to low levels of the 2\SSst\ state, taken from the list published in the supplementary materials to Ref.~\cite{Stein3}. Unfortunately, according to the selection rules specific to PLS \cite{Ferber}, when labelling levels via P or R lines (the only ones allowed in the $^1\Sigma - ^1\Sigma$ transition), the polarisation spectra were also limited to P and R lines only, irrespectively of symmetry of the upper state. This was an unfortunate drawback which prevented us from direct distinguishing between $\Sigma \leftarrow \Sigma$ and $\Pi \leftarrow \Sigma$ transitions in the recorded spectra. However, in the frequency range covered by our ring dye laser and at the same time by the available spectral data there was no other choice of labelling transitions.

The Sr$_2$ molecules were produced in a three section heat-pipe oven \cite{Ciamei2} of 1~m length, with the inner surface of the pipe lined with three separate pieces of steel mesh. The oven was filled with 15~g of strontium and 1.5~g of metallic magnesium. As strontium and magnesium form an alloy with substantially lower melting point than its constituents \cite{Nayeb}, this experimental trick ensured a proper circulation of the metal inside the heat-pipe and, in particular, prevented condensation of solid strontium in the outer zones of the oven. The central part of the heat-pipe (about 20~cm in length) was heated to 1020$^{\circ}$C, while external parts were maintained at 720$^{\circ}$C and it was operated with argon as a buffer gas at a pressure of 15~Torr. 

\section{Experimental results}
\label{Exp_res}

The present experiment provided frequencies of 2556 spectral lines, all of them originating from the ground state X$^1\Sigma^{+}_{\mathrm{g}}$ of the $^{88}$Sr$_2$ isotopologue, measured in the range $13000-15500$~\rcm. The line frequencies were converted to energies of the corresponding upper levels using highly precise molecular constants of the ground state \cite{Stein2}. In addition, we had to our disposal 287 levels of $^{88}$Sr$_2$ listed in the supplementary materials to the paper by Stein \textit{et al.} \cite{Stein3}. The data by Stein cover the range of vibrational levels $v'=1-20$ but the majority of levels corresponds to $v' \le 10$. The energies of rovibrational levels in the range $1 \le v' \le 10$ and $J'<220$ were represented in Ref.~\cite{Stein3} by a set of as many as 17 Dunham coefficients, most of them being merely fitting parameters and having no physical meaning. Stein \textit{et al.} claimed that attempts to incorporate higher vibrational levels into the fit failed, probably due to perturbation of these levels. However a scarcity of data in this range made further analysis futile. As the observations from the present experiment, providing a vast majority of the combined data, cover energy levels with rotational quantum numbers $J'$ below 150, for further discussion we have discarded higher rotational levels from the database of Ref.~\cite{Stein3}. Our analysis of the available data consists in two parts.

For the energy levels $1 \le v' \le 10$ we confirm that they can be described by a Dunham series in a form

\begin{equation}
	T(v',J')=T_e+\sum_{m,n}Y_{mn}(v'+0.5)^{m}[J'(J'+1)]^{n} \mbox{ .}
	\label{1}
\end{equation}

\noindent In this range we added 291 levels measured in the present experiment to 133 levels listed by Stein \textit{et al.} Fits with various numbers of parameters have shown that only 8 Dunham coefficients suffice to reproduce experimental level energies with an rms error of 0.08~\rcm, consistent with our experimental accuracy. A considerable reduction of the number of parameters is apparently connected with exclusion from the fit relatively few levels with $J' \ge 150$ (37 levels) present in the data of Stein \textit{et al.} The Dunham coefficients, rounded to the minimum number of significant figures by a procedure suggested by Le Roy \cite{LeRoy}, are presented in Table~\ref{table:Tab1}. We found that even addition of the next vibrational level, $v'=11$, makes description by a Dunham series impossible, as then it would require many more Dunham coefficients, still providing an rms error above 0.1~\rcm.

\begin{table}
	\centering
	\caption{The Dunham coefficients that describe the 1$^1\Sigma^{+}_{\mathrm{u}}$
		state of $^{88}$Sr$_2$ in the range $1 \leq v' \leq 10$, 
		$J'<150$ wirh rms error 0.08~\rcm. The numbers in parentheses correspond to one standard deviation in units of the last digits. Values of the coefficients are in \rcm.
	} \label{table:Tab1}\vspace*{0.5cm}
	\begin{tabular}{cccc}
		\hline Coefficient & Value \\
		\hline \
		$T_e$	&	12795.70(4) \\
		$Y_{10}$  &	80.314(23) \\
		$Y_{20}$  &	-0.1944(38) \\
		$Y_{30}$   &	-0.76(2)$\times 10^{-2}$ \\
		$Y_{01}$   &	0.0247559(39) \\
		$Y_{11}$   &	-0.8935(12)$\times 10^{-4}$ \\
		$Y_{21}$   &	-0.226(8)$\times 10^{-5}$ \\
		$Y_{02}$   &	-0.103(7)$\times 10^{-7}$ \\
		\hline \
	\end{tabular}
\end{table}

\begin{figure}[!h]
	\includegraphics[width=.95\linewidth]{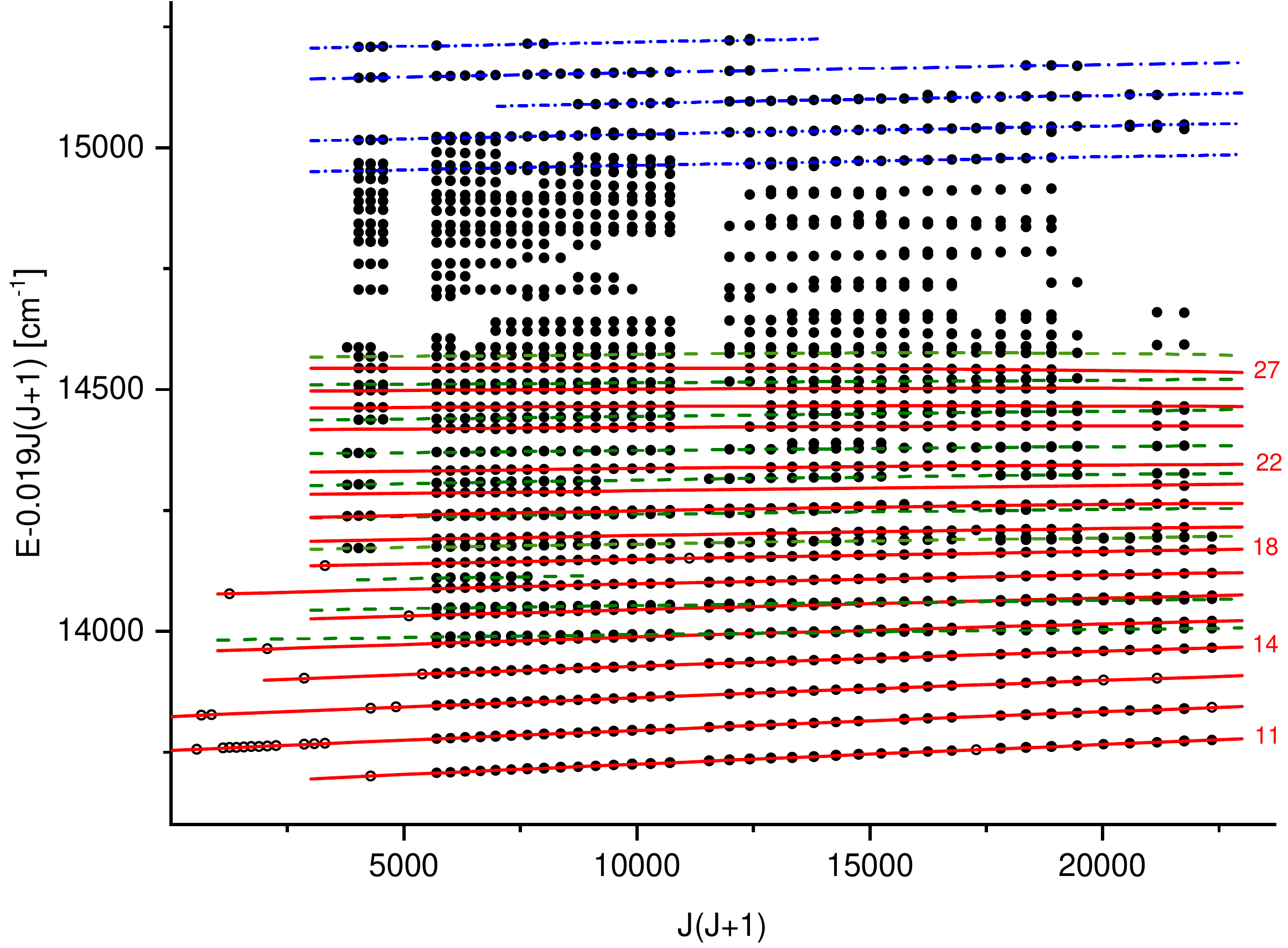}
	\caption{Reduced term values $E-0.019 \times J(J+1)$ [\rcm] of the observed rovibrational levels in the 1\SSst\ $\sim$ 2\TSst\ $\sim$ 1\TPst\ $\sim$ 2\TPst\ complex in Sr$_2$ (black dots) plotted against $J(J+1)$. The term values calculated from the band parameters of the 1\SSst\ state listed in Table~\ref{table:Tab2} are represented by red solid lines and the vibrational numbering is shown for some of them. Green dashed lines correspond to 'perturber~1' (Table~\ref{table:Tab3}) and blue dash-dot lines to 'perturber~2' (Table~\ref{table:Tab4}).}
	\label{Fig2}
\end{figure}

As indicated above, the energies of rovibrational levels of the 1\SSst\ state above $v'=10$ display rather erratic behaviour and starting from $v'=15$ levels of more than one electronic state become visible. All the observed levels in this range are displayed in Figure~\ref{Fig2}. Several of them can be arranged into bands apparently corresponding to separate, albeit strongly perturbed, vibrational levels. It must be stressed that for a vast majority of the observed rovibrational levels we are certain about their attribution to given $J'$ quantum numbers and to the main isotopologue $^{88}$Sr$_2$ because they were intentionally excited more than once, via different transitions from at least two initial ground state levels, what confirmed their assignment. Moreover, in most spectra only a single ($v''$, $J''$) level in the ground state was labelled due to a small linewidth of the probe laser (ca 1~MHz).

Despite the observed line shifts amounting in some cases to few \rcm, we tried to determine approximate band constants of several levels, after discarding rather arbitrarily the obvious outliers, by fitting level energies to the formula

\begin{equation}
	E(v',J') = T_v + B_v[J'(J'+1)] - D_v[J'(J'+1)]^2 \mbox{ .}
	\label{2}
\end{equation}

\noindent The results are listed in Tables~\ref{table:Tab2}, \ref{table:Tab3}, \ref{table:Tab4} and displayed by lines in Figure~\ref{Fig2}. We were able to follow systematically vibrational levels of the 1\SSst\ state up to $v'=27$, in spite of their irregular behaviour (red lines in Figure~\ref{Fig2}). This left few other bands which must be attributed to another electronic state ('perturber~1' -- green lines in Figure~\ref{Fig2}). Beyond $v'=27$ of the 1\SSst\ state the pattern of rovibrational levels becomes incomprehensible but above an excitation energy $E \approx 14900$~\rcm\ another system of bands emerges ('perturber~2' -- blue lines in Figure~\ref{Fig2}) belonging to an electronic state with different characteristics (cf. different slopes of the blue lines versus the red and green ones, indicating larger values of vibrational constants $B_v$ of the uppermost state).

\begin{table}
	\centering
	\caption{Approximate band constants (all values in \rcm) of vibrational levels $v'=11-27$ of the 1$^1\Sigma^{+}_{\mathrm{u}}$ state (with no deperturbation attempted).}
	\label{table:Tab2} \vspace*{0.5cm}
	\begin{tabular}{cccccccc}
		\hline
		& $v'$ &  $T_v$ & $B_v$& $D_v$ $\times10^8$ & rms & range of $J'$ \\
		\hline \
		&11	&13681.66	&0.023464	&1.31		&0.07	&$65-149$ \\
		&12	&13753.65	&0.023283	&1.46		&0.04	&$23-149$ \\
		&13	&13823.81	&0.023084	&1.80		&0.04	&$25-145$ \\
		&14	&13891.51	&0.022832	&2.32		&0.05	&$53-149$ \\
		&15	&13956.14	&0.022432	&2.42		&0.08	&$45-149$ \\
		&16	&14017.37	&0.021914	&1.78		&0.06	&$71-149$ \\
		&17	&14074.47	&0.021593	&2.38		&0.06	&$35-149$ \\
		&18	&14129.15	&0.021134	&1.64		&0.04	&$57-149$ \\
		&19	&14180.14	&0.020985	&1.86		&0.08	&$75-147$ \\
		&20	&14228.77	&0.021433	&3.78		&0.08	&$61-147$ \\
		&21	&14280.05	&0.020059	&---		&0.03	&$75-95$ \\
		&22	&14325.14	&0.020337	&1.98		&0.08	&$75-139$ \\
		&23	&---		&---		&---  	    & ---	&---	  \\
		&24	&14414.29	&0.019939	&2.08		&0.07	&$77-139$ \\
		&25	&14458.99	&0.019940	&2.97		&0.08	&$65-145$ \\
		&26	&14508.10	&0.019481	&0.14		&0.12	&$81-139$ \\
		&27	&14543.83	&0.019075	&1.36		&0.04	&$65-137$ \\

		\hline
	\end{tabular}
\end{table} 

\begin{table}
	\centering
	\caption{Approximate band parameters (all values in \rcm) of vibrational levels of the electronic state referred to as 'perturber~1'. The absolute vibrational numbering is unknown.}
	\label{table:Tab3} \vspace*{0.5cm}
	\begin{tabular}{cccccccc}
		\hline
		& $v'$ &  $T_v$ & $B_v$& $D_v$ $\times10^8$ & rms & range of $J'$ \\
		\hline \
		
		&n	&13980.74	&0.020312	&0.82		&0.05	&$75-149$ \\
		&n+1	&14039.99	&0.020472	&1.33		&0.07	&$75-149$ \\
		&n+2	&14099.64	&0.021116	&5.04		&0.06	&$75-87$ \\
		&n+3	&14165.24	&0.020682	&1.45		&0.04	&$75-149$ \\
		&n+4	&14232.05	&0.020015	&0.21		&0.04	&$75-135$ \\
		&n+5	&14296.62	&0.020852	&2.37		&0.12	&$63-135$ \\
		&n+6	&14364.31	&0.020055	&0.85		&0.08	&$61-147$ \\
		&n+7	&14433.28	&0.020145	&0.19		&0.12	&$77-137$ \\
		&n+8	&14494.29	&0.019914	&2.51		&0.06	&$63-131$ \\
		\hline
	\end{tabular}
\end{table} 

\begin{table}
	\centering
	\caption{Approximate band parameters (all values in \rcm) of vibrational levels of the electronic state referred to as 'perturber~2'. The absolute vibrational numbering is unknown.}
	\label{table:Tab4} \vspace*{0.5cm}
	\begin{tabular}{cccccccc}
		\hline
		& $v'$ &  $T_v$ & $B_v$& $D_v$ $\times10^8$ & rms & range of $J'$ \\
		\hline \
		
		&p	&14944.49	&0.021006	&0.85		&0.08	&$63-137$ \\
		&p+1	&15008.01	&0.021072	&1.02		&0.20	&$63-147$ \\
		&p+2	&15071.62	&0.021202   &1.80		&0.14	&$93-137$  \\
		&p+3	&15136.38	&0.021074	&1.53		&0.08	&$63-111$ \\
		&p+4	&15200.54	&0.020924	&1.08		&0.13	&$63-109$ \\
		
		\hline
	\end{tabular}
\end{table}

It must be noted that Leung \textit{et al.} \cite{Leung} claimed to measure energies of 37 lowest vibrational levels of the 1\SSst\ state but the measurements, as usually in the ultracold physics experiments, were limited to only one rotational level $J'=1$ for each vibrational quantum number. We find that energies of these levels agree with values calculated from parameters of Table~\ref{table:Tab2} within $\pm 0.2$~\rcm\ for $v'=11-18$, with discrepancies growing irregularly up to $\pm 2$~\rcm\ for strongly perturbed levels $v'=19-27$. This agreement can be considered good taking into account long extrapolation from the range of rotational quantum numbers explored in our experiment to $J'=1$ as well as somewhat arbitrary choice of 'unperturbed' levels used in our fits.

\section{Discussion}
\label{Dis}

Because of large discrepancies in potential energy curves derived by different authors and methods, the discussion in this section can be only of qualitative nature. We believe that the oldest calculations by Boutassetta \textit{et al.} \cite{Allouche,Allouche2} (Figure~\ref{Fig1}) provide a plausible explanation of the experimental findings. According to Figure~\ref{Fig1}, in a simplistic picture, the 1\SSst\ state should be free of perturbation by any other electronic state up to $E \approx 13650$~\rcm\ where the origin of the 2\TSst\ state is located. However, a direct perturbation 1\SSst\ $\sim$ 2\TSst\ is not possible \cite{Kovacs} and it has to be mediated via mixing of the 2\TSst\ and 1\TPst\ states. Our experimental observations suggest that this is the 1\TPst\ state which becomes visible in the spectra as 'perturber~1' (note smaller $B_v$ values corresponding to levels marked in green in Figure~\ref{Fig2}, which indicate larger $R_v$ values than in the 1\SSst\ state). With increasing excitation energy, the inner turning points of the 1\TPst\ and 2\TSst\ potentials gradually draw closer and mutual perturbation of both triplet states makes the pattern of vibrational levels even more complicated (cf. the region of reduced energies $14500-15000$~\rcm\ in Figure~\ref{Fig2}). Finally, at approximately 14900~\rcm\ another state emerges, namely 2\TPst, which can be visible in the spectra due to a direct 1\SSst\ $\sim$ 2\TPst\ perturbation ('perturber~2').

\begin{figure}[!h]
	\includegraphics[width=0.95\linewidth]{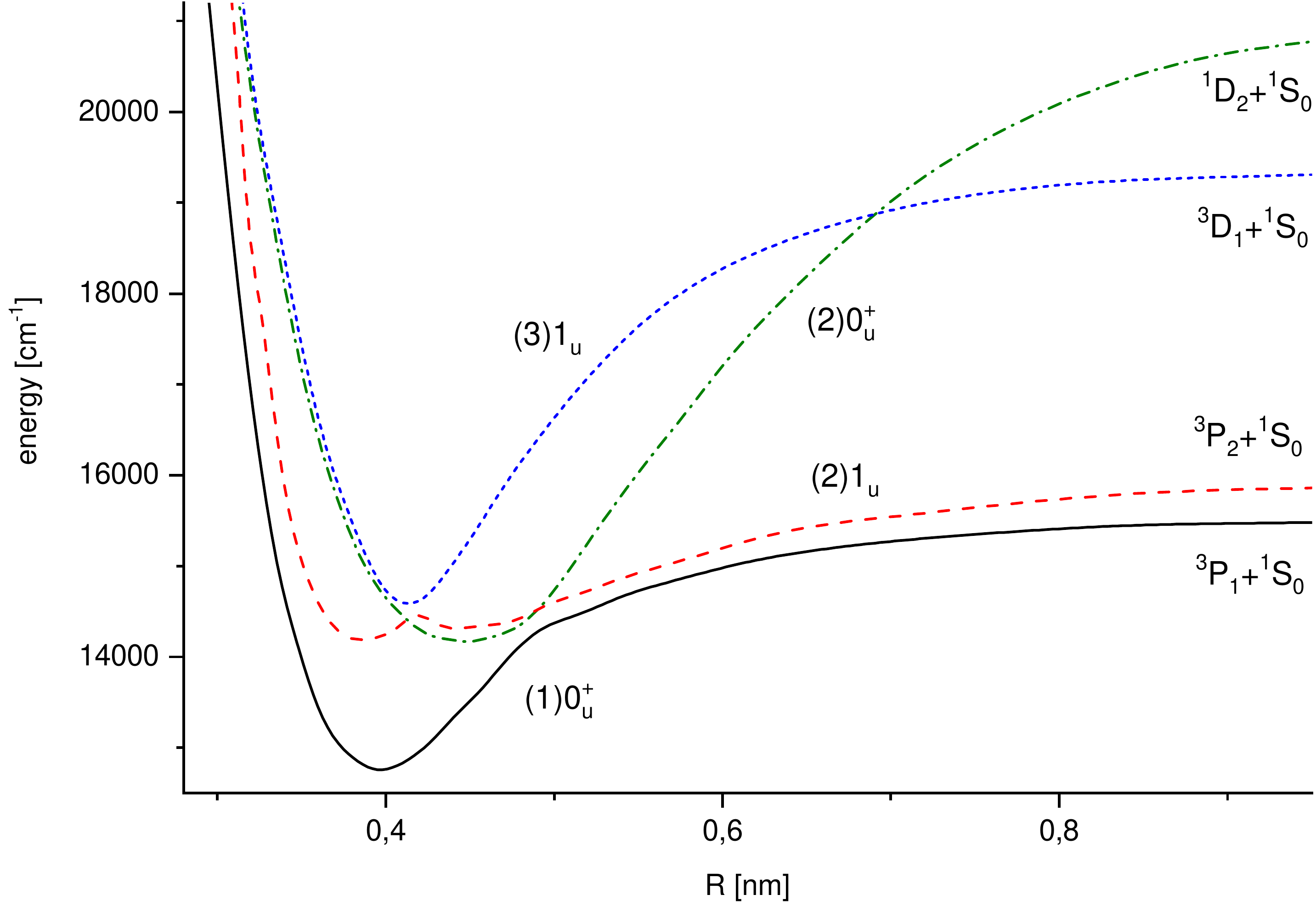}
	\caption{Potential energy curves of Sr$_2$ relevant to the present work, calculated in Hund's case (c) by Skomorowski \textit{et al.} \cite{Skomorowski2}. (As Ref.~\cite{Skomorowski2} does not provide numerical results referring to the potential curves, this Figure is an illustrative redrawing of their Fig.3.)}
	\label{Fig3}
\end{figure}

An interesting, alternative explanation of the experimental observations may be provided by potential curves of Sr$_2$ calculated in Hund's case (c) by Skomorowski \textit{et al.} \cite{Skomorowski2} (Figure~\ref{Fig3}). Note that in absence of singlet--triplet symmetry of electronic states all transitions between them should be visible, as observed in our experiment. In Hund's case (c) the 1\SSst\ state corresponds to (1)$0^+_u$ and then the sequence of states coming into play when excitation energy increases is similar as in case (a), here 'perturber~1' being the (2)$0^+_u$ state and 'perturber~2' -- the (3)$1_u$ state. Skomorowski \textit{et al.} claim very good agreement between the low part of their (1)$0^+_u$ state potential and the experimental one constructed by Stein \textit{et al.} \cite{Stein3} but again we find that theoretical results describe our findings only qualitatively, particularly that no numerical data accompany Ref.~\cite{Skomorowski2}.

The above analysis shows clearly that a rigorous deperturbation of the coupled 1\SSst\ $\sim$ 2\TSst\ $\sim$ 1\TPst\ $\sim$ 2\TPst\ complex of states is necessary to explain the experimental results quantitatively. This is a serious numerical challenge, but first of all it would require reliable theoretical potentials as starting points. In the lack of them, in addition to qualitative analyses we make available the full list of the experimentally determined energies of rovibrational levels as the supplementary material accompanying this paper.

\section{Acknowledgements}

Partial financial support from the National Science Centre Poland (Grant No. 2021/43/B/ST4/03326) is gratefully acknowledged.

\section{Appendix A. Supplementary data}

Supplementary data associated with this article can be found online at the address http://dimer.ifpan.edu.pl.

\end{document}